\documentclass[twocolumn]{aastex62}
\pdfoutput=1 
\usepackage{natbib}
\usepackage{amsmath}
\usepackage{graphicx} 
\usepackage{amssymb}
\usepackage{pgffor}

\shorttitle{CO($J = 3-2$) in ACT\,J0209}
\shortauthors{Rivera et al.}


\begin{document}

\title{The Atacama Cosmology Telescope: \\ CO($J = 3-2$) mapping and lens
  modeling of an ACT-selected dusty star-forming galaxy}


\correspondingauthor{Jesus Rivera}
\email{jrivera@physics.rutgers.edu}

\author[0000-0003-3191-5193]{Jesus Rivera} 
\affil{Department of Physics and Astronomy, Rutgers, the State 
  University of New Jersey, 136 Frelinghuysen Road, Piscataway, NJ 08854-8019,
  USA}

\author[0000-0002-7892-396X]{Andrew J. Baker}
\affiliation{Department of Physics and Astronomy, Rutgers, the State 
  University of New Jersey, 136 Frelinghuysen Road, Piscataway, NJ 08854-8019,
  USA}
  
\author{Patricio A. Gallardo} 
\affiliation{Department of Physics, Cornell University, 109 Clark Hall, Ithaca, NY
	14853-2501, USA}

\author{Megan B. Gralla}
\affiliation{Steward Observatory, University of Arizona, 933 North Cherry
  Avenue, Tucson, AZ 85721, USA}

\author[0000-0001-6159-9174]{Andrew I. Harris}
\affiliation{Department of Astronomy, University of Maryland, College Park,
  MD 20742-2421, USA}

\author[0000-0001-7109-0099]{Kevin M. Huffenberger}
\affiliation{Department of Physics, Florida State University, Tallahassee, FL
  32306, USA}

\author[0000-0002-8816-6800]{John P. Hughes}
\affiliation{Department of Physics and Astronomy, Rutgers, the State 
  University of New Jersey, 136 Frelinghuysen Road, Piscataway, NJ 08854-8019,
  USA}

\author[0000-0001-6812-2467]{Charles R. Keeton}
\affiliation{Department of Physics and Astronomy, Rutgers, the State 
  University of New Jersey, 136 Frelinghuysen Road, Piscataway, NJ 08854-8019,
  USA}

\author{Carlos H. L\'opez-Caraballo}
\affiliation{Instituto de Astrof\'isica and Centro de Astro-Ingenier\'ia, 
	Facultad de F\'isica, Pontificia Universidad Cat\'olica de Chile, Av. 
	Vicu\~na Mackenna 4860, 7820436 Macul, Santiago, Chile}

\author{Tobias A. Marriage}
\affiliation{Department of Physics and Astronomy, The Johns Hopkins University,
  3400 North Charles Street, Baltimore, MD 21218-2686, USA}

\author[0000-0001-6541-9265]{Bruce Partridge}
\affiliation{Department of Physics and Astronomy, Haverford College,
  Haverford, PA 19041, USA}

\author[0000-0001-6903-5074]{Jonathan L. Sievers}
\affiliation{University of KwaZulu-Natal, Durban, South Africa National
  Institute for Theoretical Physics, Durban 4000, KwaZulu-Natal, South Africa}
\affiliation{Department of Physics, McGill University, Montreal, Quebec, H3A
  2T8, Canada}

\author{Amitpal S. Tagore}
\affiliation{Jodrell Bank Centre for Astrophysics, The University of Manchester,
  Manchester, M13 9PL, UK}

\author[0000-0003-4793-7880]{Fabian Walter}
\affiliation{Max-Planck-Institut f\"ur Astronomie, K\"onigstuhl 17, D-69117,
  Heidelberg, Germany}

\author{Axel Wei{\ss}}
\affiliation{Max-Planck-Institut f\"ur Radioastronomie, Auf dem H\"ugel 69,
  D-53121 Bonn, Germany}

\author[0000-0002-7567-4451]{Edward J. Wollack}
\affiliation{NASA/Goddard Space Flight Center, Greenbelt, MD 20771, USA}


\begin{abstract}
We report Northern Extended Millimeter Array (NOEMA) CO($J = 3 - 2$)
observations of the dusty star-forming galaxy ACT-S\,J020941+001557 at
$z = 2.5528$, which was detected as an unresolved source in the Atacama
Cosmology Telescope (ACT) equatorial survey.  Our spatially resolved spectral
line data support the derivation of a gravitational lens model from 37
independent velocity channel maps using a pixel-based algorithm, from which
we infer a velocity-dependent magnification factor $\mu \approx 7-22$ with
a luminosity-weighted mean $\left<\mu\right>\approx 13$.  The resulting
source-plane reconstruction is consistent with a rotating disk, although other
scenarios cannot be ruled out by our data. After correction for lensing, we
derive a line luminosity $L^{\prime}_{\rm CO(3-2)}= (5.53\pm 0.69) \times
10^{10}\,{\rm \,K\,km\,s^{-1}\,pc^{2}}$, a cold gas mass $M_{{\rm gas}}= (3.86
\pm 0.33) \times 10^{10}\,M_{\odot}$, a dynamical mass $M_{\rm dyn}\,{\rm
sin}^2\,i = 3.9^{+1.8}_{-1.5} \times 10^{10}\,M_{\odot}$, and a gas mass fraction 
$f_{\rm gas}\,{\rm csc}^2\,i = 1.0^{+0.8}_{-0.4}$. The line brightness temperature
ratio of $r_{3,1}\approx 1.6$ relative to a Green Bank Telescope CO($J=1-0$) detection
may be elevated by a combination of external heating of molecular clouds,
differential lensing, and/or pointing errors. 
\end{abstract}

\section{Introduction}
The characterization of dusty star-forming galaxies (DSFGs) at high redshift 
entered a new era with the discovery of submillimeter galaxies 
\citep{smail97,hughes98,barger98}. These pioneering studies and their
 successors revealed a large, previously unknown population 
of star-forming systems in which dust absorbs nearly all of the UV and optical 
emission radiated from stars and active galactic nuclei (AGN) 
and re-emits it in the far-infrared/submillimeter regime.  DSFGs, now
defined more broadly to include systems selected at wavelengths other than the
submillimeter, play a critical role in the history of galaxy formation
and evolution. They are substantial contributors to the cosmic star formation 
history and the likely progenitors of  nearby elliptical galaxies 
\citep{blain2002,casey2014}. However, our understanding of these 
galaxies is still limited by our ability to detect and to study them 
in detail; recent efforts by many teams have focused on this goal. 

Wide-field surveys that map areas greater than $100$\,deg$^{2}$ with
sufficient angular resolution have been
used to discover gravitationally lensed DSFGs at far-IR and (sub)millimeter 
wavelengths with \textit{Herschel} \citep{negrello2010,wardlow2013,nayyeri2016},
the South Pole Telescope \citep{vieira2010,mocanu2013}, the Atacama
Cosmology Telescope \citep[ACT;][]{marsden2014}, and \textit{Planck}
\citep{canameras2015}.  High flux density tails on observed number count
distributions have proved to be a mixture of strongly lensed DSFGs, clusters
of DSFGs, and ``trainwreck'' systems in which complex mergers of multiple,
modestly lensed DSFGs are blended into a single bright source
\citep{riechers2011a,ivison2013,fu2013,miller2018}. Strong lensing enables the
study of lens-plane mass distributions (including substructure), while
allowing the observation of distant, intrinsically faint galaxies in the source
plane that would otherwise be too dim to detect.  In order to probe the physical
properties of lensed DSFGs, we can observe rotational emission lines (most
notably of CO) that trace molecular gas and star-forming material. 

In this article, we present CO$(J$\,=\,$3-2)$  observations of the $z=2.5528$
DSFG ACT-S\,J020941+001557 (ACT\,J0209\footnote{\citet{su2017} refer to the
  source as ACT-S\,J0210+0016.}) with the Northern Extended Millimeter
Array (NOEMA).  Section 2 describes previous and new observations; Section 3
analyzes the spectral line properties, gravitational lensing, and source-plane
gas morphology of ACT\,J0209; and Section 4 discusses our conclusions. All
calculations assume a flat cosmology with
$H_{\rm 0} = 71\,{\rm km\,s^{-1}\,Mpc^{-1}}$, 
$\Omega_M = 0.27$, and $\Omega_{\Lambda} = 0.73$.

\section{Observations} \label{s-obs}

\subsection{Previous observations}

ACT\,J0209 was detected with ACT \citep{fowler2007,swetz2011} in a
$470$\,deg$^{2}$ equatorial survey at 148, 218, and 277\,GHz \citep[Gralla et
al. 2019, in preparation;][]{su2017}.  DSFG candidates were selected based on
their 218\,GHz flux densities ($>18$\,mJy) and consistency of their 148$-$218\,GHz spectral indices with thermal dust emission. The ACT equatorial field was purposely
designed to overlap the Sloan Digital Sky Survey (SDSS) Stripe 82 footprint
\citep{abazajian2009} and 1.4\,GHz imaging by the Very Large Array
\citep{becker1995} to facilitate the identification of optical and 
radio counterparts.  Our initial followup of the ACT detection of ACT\,J0209
($S_{218} = 69.2 \pm 2.7\,{\rm mJy}$) was reported in \citet{su2017} and 
included CO($J = 1 - 0$) observations with the Robert C. Byrd Green Bank 
Telescope (GBT) and subsequent low-resolution CO($J = 3 - 2$) imaging with the 
Combined Array for Research in Millimeter-wave Astronomy (CARMA).  The latter 
dataset confirmed a DSFG redshift $z = 2.5528$ in excess of the SDSS redshift 
for its optical counterpart ($z_{\rm lens} = 0.202$), highly suggestive of 
lensing. The source's redshift, lensed status, and radio-loud AGN were first 
reported by \citet{geach2015}, who designated it as 9io9 and dubbed it the 
``Red Radio Ring'' following its identification in a citizen science project to
find lenses. Shortly before the submission of this manuscript, \citet{geach2018} reported an analysis of $\sim 0.25^{\prime\prime}$ resolution Atacama Large
Millimeter/submillimeter Array (ALMA) observations of ACT\,J0209 
in the CO($J = 4-3$), \ion{C}{1}($J = 1-0$), and CN($N = 4-3$) lines, which we discuss further below.  For our assumed cosmology, $D_A = 679\,{\rm Mpc}$ (1.68\,Gpc) and $D_L = 982\,{\rm Mpc}$
(21.2\,Gpc) at $z_{\rm lens} = 0.202$ ($z = 2.5528$).   

\subsection{New NOEMA observations}

We observed ACT\,J0209 with NOEMA on 2016 December 28 and 2017 January 08
(program W16DX; PI, A. Wei\ss), targeting a J2000 position of
$\alpha$=02h\,09m\,40.80s and
$\delta$=+00h\,15m\,57.60s.  The array's eight 15\,m diameter antennas were
deployed in the A configuration with a longest baseline of 760\,m.  Receivers
were tuned to 97.33\,GHz to detect CO($J = 3-2$) emission at a redshift
$z = 2.5528$. The total on-source integration time after combining both days
and flagging bad data was 11.3\,hr. We reduced the NOEMA data using the
Institut de Radioastronomie Millim\'etrique (IRAM) GILDAS software
\citep{gildas2013}.  The sensitivity achieved for robust
weighting and deconvolution with the CLEAN algorithm is 0.5\,mJy\,beam$^{-1}$
in a 20\,km\,s$^{-1}$ channel. We used 3C454.3 as a bandpass calibrator, 0221+067 and 0215+015 as
complex gain calibrators, and MWC349 as a flux calibrator, for which we adopted a flux density of 1.14\,Jy at 97.3\,GHz with $\sim 10\%$ uncertainty \citep[e.g.,][]{trippe2012}. 

\begin{figure}[ht!]
\centering
\includegraphics[scale=0.36]{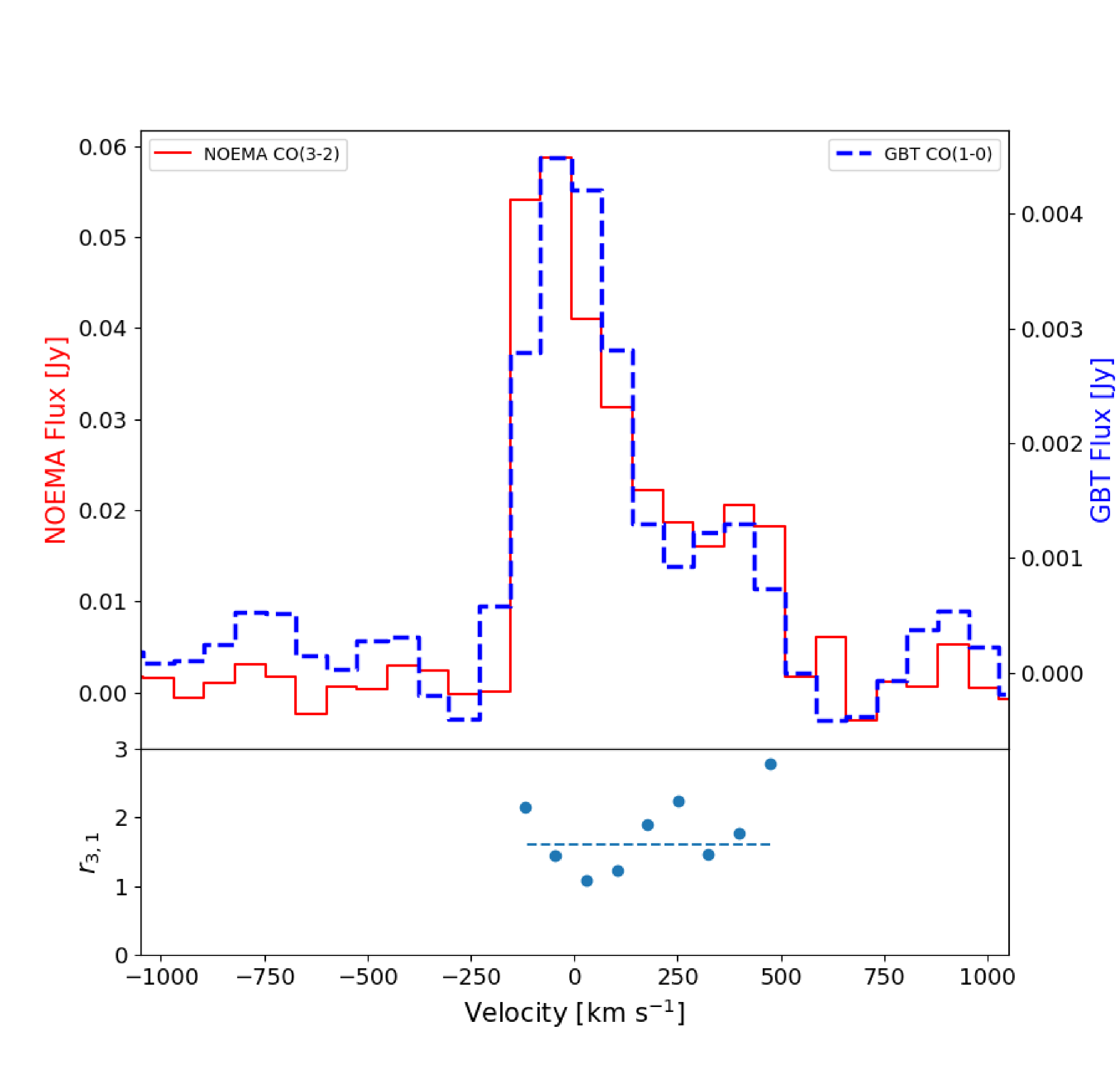}
\caption{Upper panel: NOEMA CO($J = 3-2$) spectrum of ACT\,J0209 (red)
  overlaid with the GBT CO($J = 1-0$) spectrum (dashed blue). The NOEMA data
  have been rebinned to match the coarser resolution of the GBT data.
  Integrated line fluxes are summed over FWZI\,=\,720\,km\,s$^{-1}$.  Lower
  panel: $r_{3,1}$ ratio on a per channel basis.  The luminosity-weighted
  global mean $\left<r_{3,1}\right> \approx 1.6$ is indicated as a dashed line.}
\end{figure}

\begin{deluxetable*}{cccccccc}
	\tablecolumns{8}
	\tablewidth{0pt}
        \tablecaption{Lens model parameters. The ``best'' model yielded the single lowest $\chi^{2}$ model in our MCMC runs and is used for all analyses of reconstructed source-plane emission in this paper. The ``MCMC 
        median'' model shows the median value of each parameter with uncertainties 
        corresponding to the 16th and 84th percentile ranges derived from the MCMC 
        sampling; the small uncertainties in some positional parameters reflect priors imposed based on {\it HST} imaging. Due to the complex multidimensional parameter space, not every ``best'' model parameter falls within the 16th--84th percentile range for the corresponding ``MCMC median'' distribution.} 
	\tablehead{
		\colhead{} &
		\colhead{$b$} &
 		\colhead{$\Delta \alpha$} &  
		\colhead{$\Delta \delta$} &
		\colhead{} &
		\colhead{} &
		\colhead{} &
		\colhead{} \\
		\colhead{} &
		\colhead{(arcsec)} &
		\colhead{(arcsec)} &
		\colhead{(arcsec)} &
		\colhead{$e_c$} &
		\colhead{$e_s$} &
		\colhead{$\gamma_c$} &
		\colhead{$\gamma_s$}}
	\decimals
	\startdata
	Best \\
    component 1 & $2.444$ & $-0.005$ & $-0.249$ & $-0.206$ & $-0.087$ & --- & --- \\
	component 2 & $0.103$ & $-0.609$ & $2.192$ & --- & --- & --- & --- \\
    external shear & --- & --- & --- & --- & --- & $-0.020$  & $0.015$ \\ 
    \tableline
    MCMC median \\
	component 1 & $2.436^{+0.016}_{-0.010}$ & $-0.005^{+0.000}_{-0.001}$ & $-0.241^{+0.009}_{-0.007}$ & $-0.205^{+0.028}_{-0.019}$ & $-0.106^{+0.019}_{-0.015}$ & --- & --- \\
	component 2 & $0.107^{+0.007}_{-0.005}$ & $-0.570^{+0.048}_{-0.030}$ & $2.193^{+0.052}_{-0.030}$ & --- & --- & --- & --- \\
    external shear & --- & --- & --- & --- & --- & $-0.0212^{+0.015}_{-0.011}$  & $0.0255^{+0.006}_{-0.007}$ \\
	\enddata
	\tablecomments{Offsets in right ascension and declination
          ($\Delta \alpha$, $\Delta \delta$) are
defined relative to the NOEMA pointing center.  The singular isothermal
ellipsoid has projected surface density $\kappa = b/2\zeta$ as implemented
in {\it gravlens} for coordinate $\zeta \equiv [(1-\epsilon)x^2 +
(1+\epsilon)y^2]^{1/2}$ with $q^2 \equiv (1-e)^2 \equiv
(1-\epsilon)/(1+\epsilon)$.  The ellipsoid's ellipticity $e$ and the position
angle (east of north) of its major axis $\theta_e$ are combined into the fit
parameters $e_c \equiv e \cos 2\theta_e$ and $e_s \equiv e \sin 2\theta_e$; the
external shear $\gamma$ and its position angle $\theta_\gamma$ are likewise
combined into fit parameters $\gamma_c$ and $\gamma_s$.
}
\end{deluxetable*}

\section{Results}

\subsection{Integrated line properties}

We detect and spatially resolve the CO($J = 3-2$) emission from ACT\,J0209; including a 10\% uncertainty in the flux scale,
the observed line flux is $F_{\rm CO(3-2)} = 20.8 \pm 0.5\,(\pm 2.1)\,{\rm Jy\,km\,s^{-1}}$, with the line extending
over a full width at zero intensity (FWZI) of 720\,km\,s$^{-1}$,
from $-190$ to $+530$\,km\,s$^{-1}$ relative to the rest frame at $z=2.5528$
(Figure 1).  Our total line flux is consistent with the value of
$18.2 \pm 2.0\,{\rm Jy\,km\,s^{-1}}$
measured with CARMA by \citet{su2017}, suggesting that the much fainter
($9.5 \pm 0.6\,{\rm Jy\,km\,s^{-1}}$) detection reported by
\citet{harrington2016} from Large Millimeter Telescope observations
may be unreliable.  Figure 1 also shows the CO($J = 1-0$) GBT data presented by 
\citet{su2017}, from which we measure a line flux $F_{\rm CO(1-0)} = 1.48 \pm
0.09\,(\pm 0.05)\,{\rm Jy\,km\,s^{-1}}$, including a term for a 3.5\% uncertainty in the reference flux density of 3C48 
\citep{perley2013}. Our NOEMA and GBT data imply a brightness temperature 
ratio of $r_{3,1} = 1.56 \pm 0.24$ (higher than inferred by \citet{su2017} on the
basis of the lower-S/N CARMA data).  
The value of $r_{3,1} \approx 1.6$ is significantly higher than seen in previous
studies of DSFGs, which yield typical ratios of $r_{3,1} \approx 0.4-1.1$ 
\citep[e.g.,][]{ivison2011,harris2012,sharon2016}, and 
cannot be explained using single-zone radiative transfer modeling alone.
\citet{su2017} argued that an elevated line ratio could be due to 
differential lensing. However, this scenario is disfavored by the broad
similarity of the velocity profiles for the GBT CO($J = 1-0$) and NOEMA
CO($J = 3-2$) spectra (Figure 1), which was not previously evident due to the
limited S/N of the CARMA data. The CO($J = 4 - 3$) velocity profile observed by \citet{geach2018}, which exhibits the same width and $\sim 3:1$ ratio between peak flux densities on the blue and red sides of the line, again offers no evidence of differential lensing of CO lines.  Our observed high $r_{3,1}$ ratio could be due in part to
imperfect GBT pointing, which would lead to an underestimated CO($J = 1-0$)
flux while preserving the line profile.  Although one recent independent GBT
CO($J = 1 - 0$) observation does imply a lower $r_{3,1} \approx 1.4$ (M.~S.
Yun, private communication), careful reassessment of our GBT data reveals no
evidence that either the overall flux scale (set by cross-calibration of the
nearby quasar 2017+0144 against 3C48) or the pointing stability during our
session can have reduced ACT\,J0209's CO($J = 1 - 0$) flux by more than
$\sim 10\%$.  Fortunately, there remains an appealing astrophysical explanation.
Values of $r_{3,1}$ as high as $1.3-1.4$ have been observed in the centers of
multiple nearby starburst galaxies \citep{dumke2001} and may result from the
external heating of molecular clouds whose CO emission is optically thick.  In
this situation --- previously invoked by multiple authors to explain observed
values $r_{2,1} > 1$ \citep[e.g.,][]{young1984,braine1992,turner1993} 
in external galaxies, and $r_{3,2} > 1$ in the Orion B molecular cloud \citep{kramer1996} ---
excitation temperature gradients can allow a higher-$J$ CO line to reach
$\tau \approx 1$ in clouds' warmer outer layers even as a lower-$J$ CO line
reaches $\tau \approx 1$ only in their cooler cores
\citep[e.g.,][]{gierens1992}. For Galactic molecular clouds, such temperature
gradients established by external UV irradiation may also help explain
anomalous combinations of CO and ${}^{13}{\rm CO}$ intensity ratios
\citep[][]{castets1990,pineda2007}.  We note that the $25.6$\,Jy\,km\,s$^{-1}$
CO($J=4-3$) flux measured for ACT\,J0209 by \citet{geach2018} implies a
brightness temperature ratio $r_{4,1} \approx 1.1$, much higher than typical
values of 0.2--0.3 in other DSFGs \citep[e.g.,][]{hainline2006,harrington2018}.
This elevated $r_{4,1}$, like our high $r_{3,1}$, can be naturally explained by
a combination of GBT mis-pointing and external heating of molecular clouds.

\subsection{Lensing reconstruction and source-plane CO properties}

Figure 2a shows the velocity-integrated line flux (i.e., zeroth moment) map, which contains an 
extended arc-like structure and a smaller counterimage.  To fit a lens model,
we follow \citet{geach2015} and assume that there are two galaxies in the
lens plane: component 1 (C1) represents the brighter galaxy, which we model as a
singular isothermal ellipsoid, and component 2 (C2) represents the fainter
galaxy, which we model as a singular isothermal sphere. Also including
external shear, we vary the 10 lens parameters shown in Table 1. We reconstruct
the source using the algorithm
described by \citet{tagore2014} and \citet{tagore2016}. Briefly,
the software treats the source as a collection of shapelets and adjusts them
to achieve the best fit to the lensed image subject to a regularization
constraint. We checked that the results are robust to different choices of
the number of shapelets.

We first fit the lens model to the zeroth moment map to obtain an initial
estimate of parameter values. We then used 37 CO($J = 3 - 2$) channel maps,
each $20\,{\rm km\,s^{-1}}$ in width, and adjusted the lens model parameters
to obtain the best joint fit to all channels.  (The sources in the different
channels are reconstructed independently.)  Our channel-map-based 
model was informed by
{\it Hubble Space Telescope} ({\it HST}) F160W imaging of the target (program
14653; PI J. Lowenthal) in two limited respects.  First, we allowed the difference between the
positions of the two lensing galaxies to diverge from their difference in the {\it HST} image by no more than $\pm 0.1^{\prime\prime}$ in right
ascension and declination, and we allowed the absolute positions of both to vary together to accommodate any {\it HST}/NOEMA astrometric inconsistency.  Second, we used the 3.25\,mag difference between
the two galaxies' integrated magnitudes, as measured by {\sc Galfit}
\citep{peng2002} fits to the {\it HST} image, to estimate a factor of $\sim 
4.5$ difference between the two mass components' Einstein radii,
on the assumption that Einstein radius
$b \propto \sigma^2 \propto L^{1/4}$ for an isothermal sphere given the
\citet{faber1976} relation.  The resulting estimate ($b = 
0.48^{\prime\prime}$) for the companion and the considerably lower value ($b = 
0.16^{\prime\prime}$) preferred by the model of \citet{geach2015} 
led us to restrict its
Einstein radius to lie in the range $0.1^{\prime\prime} - 0.5^{\prime\prime}$. 
While there are some parameter degeneracies, in particular between the mass of the
companion and the ellipticity of the main lens, we find that the model
converges to lower values of $b$ (and hence mass) regardless of what we adopt as an
initial value.  After determining our best-fit model parameters (Table 1), we
generated reconstructed channel maps (and a reconstructed moment map) for the
CO($J = 3 - 2$) data.
We then used this best-fit model as the starting point for a Markov chain 
Monte Carlo (MCMC) exploration of model parameter space, using the built-in MCMC 
function found in the \textit{gravlens} software. We started the simulation with 25 walkers 
and stopped when the solutions for each of the 10 parameters converged. The ``MCMC median'' portion of Table 1 shows the 16th--84th percentile ranges of the MCMC chains after 500 burn-in steps.

Figure 2 shows the application of our model to the CO($J = 3 - 2$) moment map.
Figure 2b shows
the model of the image plane derived from the lens reconstruction; Figure 2c
shows a (data) -- (model) residual map; and Figure 2d shows
the source-plane reconstruction.  We infer a delensed line flux
$F_{\rm CO(3-2)} = 1.60 \pm 0.20\,{\rm Jy\,km\,s^{-1}}$ and an intrinsic line
luminosity $L^{\prime}_{\rm CO(3-2)}= (5.53\pm 0.69) \times
10^{10}\,{\rm \,K\,km\,s^{-1}\,pc^{2}}$.  To estimate a cold 
molecular gas mass, we assume
the CO($J = 1 - 0$) and CO($J = 3 - 2$) lines are identically lensed (see
above), in which case the intrinsic CO($J = 1 - 0$) luminosity will be
$L^{\prime}_{\rm CO(1-0)}= (3.55 \pm 0.30) \times 10^{10}\,{\rm
  K\,km\,s^{-1}\,pc^{2}}$.  Adopting a CO-to-H$_2$ conversion factor
$\alpha_{\rm CO} = 0.8\,M_{\odot}({\rm
  K\,km\,s^{-1}\,pc^{2})^{-1}}$ appropriate for star-forming galaxies and correcting
by factor of 1.36 for helium \citet{bolatto2013}, we then infer
$M_{{\rm gas}}= (3.86 \pm 0.33) \times 10^{10}\,M_{\odot}$.

Figures 3 and 4 show channel maps of the raw data and source-plane
reconstruction, respectively.  The reconstructed channel maps reveal a clear
velocity gradient, suggesting that that the galaxy is (or at least contains) a
rotating disk.  Given the very sharp drop-offs on both sides of the velocity
profile (see below), we use the second and third most highly redshifted and
blueshifted velocity channels to estimate a circular velocity ($v_{\rm
  circ}\,{\rm sin}\,i = 340 \pm 20\,{\rm km\,s^{-1}}$) and a diameter (from
emission peaks: $2.9 \pm 0.9\,{\rm kpc}$) for the putative disk,
and thereby a dynamical mass of ${M_{\rm dyn}}\,{\rm
  sin}^2\,i = 3.9^{+1.8}_{-1.5} \times 10^{10}\,M_{\odot}$ in terms of an
inclination $i$ that we are unable to constrain.  The implied cold gas mass
fraction is thus a rather high $f_{\rm gas}\,{\rm csc}^2\,i = 1.0^{+0.8}_{-0.4}$.

The upper panel of Figure 5 shows a spectrum extracted
from the reconstructed channel maps (red) and directly compares
it to the observed spectrum (black), after scaling the two peaks of the spectra at $\sim -80$\,km\,sec$^{-1}$ to match.  Also overplotted is a spectrum extracted
from the modeled image-plane channel maps (blue), whose good
agreement with the observed spectrum validates our lens model.
The lower panel shows the inferred magnification
factor on a per channel basis, calculated by dividing the lensed spectrum 
by the de-lensed spectrum. 
Our lens reconstruction shows that ACT\,J0209 has a velocity-dependent
magnification range of $\mu\approx 7-22$ with an average of $\mu\approx 13$,
which is consistent with the result
of \citet{geach2015} that $\mu_{\rm{NIR}}\approx14$ and $\mu_{\rm{5\,GHz}}
\approx 13$. We also see that the bluer side of the line (relative to $z=2.5528$) is more 
highly magnified than the redder side. 

In order to calculate uncertainties in the magnification 
factors, we use the source brightness covariance matrix associated with our 
best model to generate $10^4$ random sources, lens those sources, and compute 
the magnification factors for each. The magnification uncertainties shown in the 
lower panel of Figure 5 correspond to the 16th and 84th percentile range of the distributions in $\mu$; these show channel-to-channel variations that (like the magnifications themselves) depend on S/N and proximity to lensing caustics.
We have also assessed the uncertainties in magnification that result from varying the {\it model} (rather than the source), using the MCMC chains described above. However, we find that model-driven uncertainties are smaller than source-driven uncertainties for all but one of the 37 channels in our dataset. Since the two types of uncertainties are not independent and thus cannot be added in quadrature, we opt in Figure 5 to show only the dominant, source-based uncertainties, plotted as 16th--84th percentile error bars.

\begin{figure}[ht!]
\centering 
\includegraphics[scale=.46]{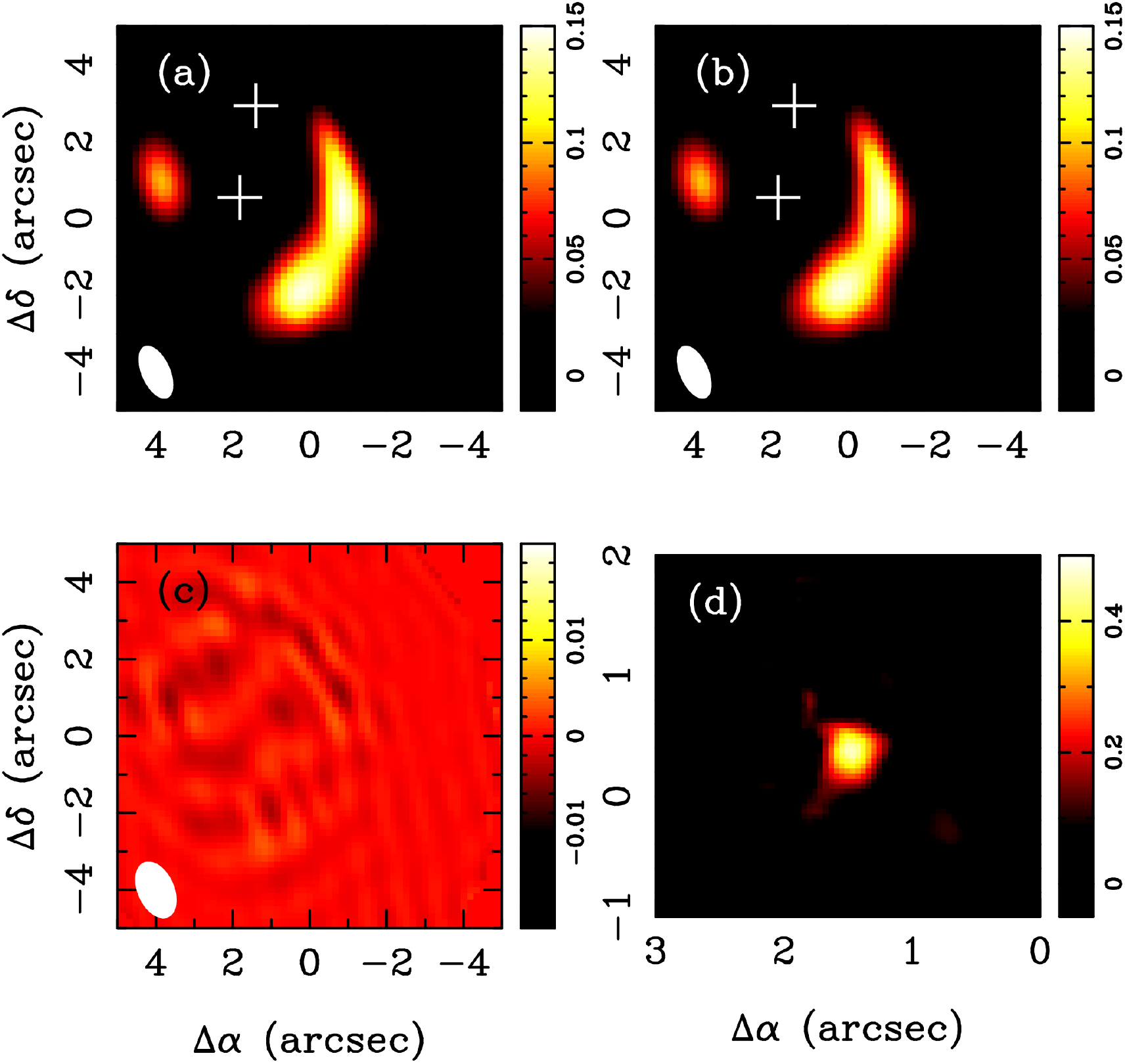}
\caption{Lens model for ACT\,J0209 applied to the CO($J = 3-2$) moment map.
  Upper left panel (a) shows observed moment map; upper right panel (b) shows
  model moment map assuming lens parameters from Table 1 and the reconstructed
  source in lower right panel (d); and lower left panel (c) shows the (data) --
  (model) residual map. The white crosses in panels (a) and (b) mark the positions
  of the two lens components. In the image-plane panels, the beam
  size shown in the lower left corners is
  $1.53^{\prime\prime}\times 0.74^{\prime\prime}$, and intensity units are Jy\,beam
  $^{-1}$\,km\,s$^{-1}$; the source-plane intensity scale differs due to its lower
  degree of effective smoothing. All panels integrate over
  ${\Delta v}= -190 \rightarrow +530\,{\rm km\,s^{-1}}$ relative to
  systemic redshift.}
\end{figure}

\begin{figure}[ht!]
\centering
\includegraphics[scale=0.40]{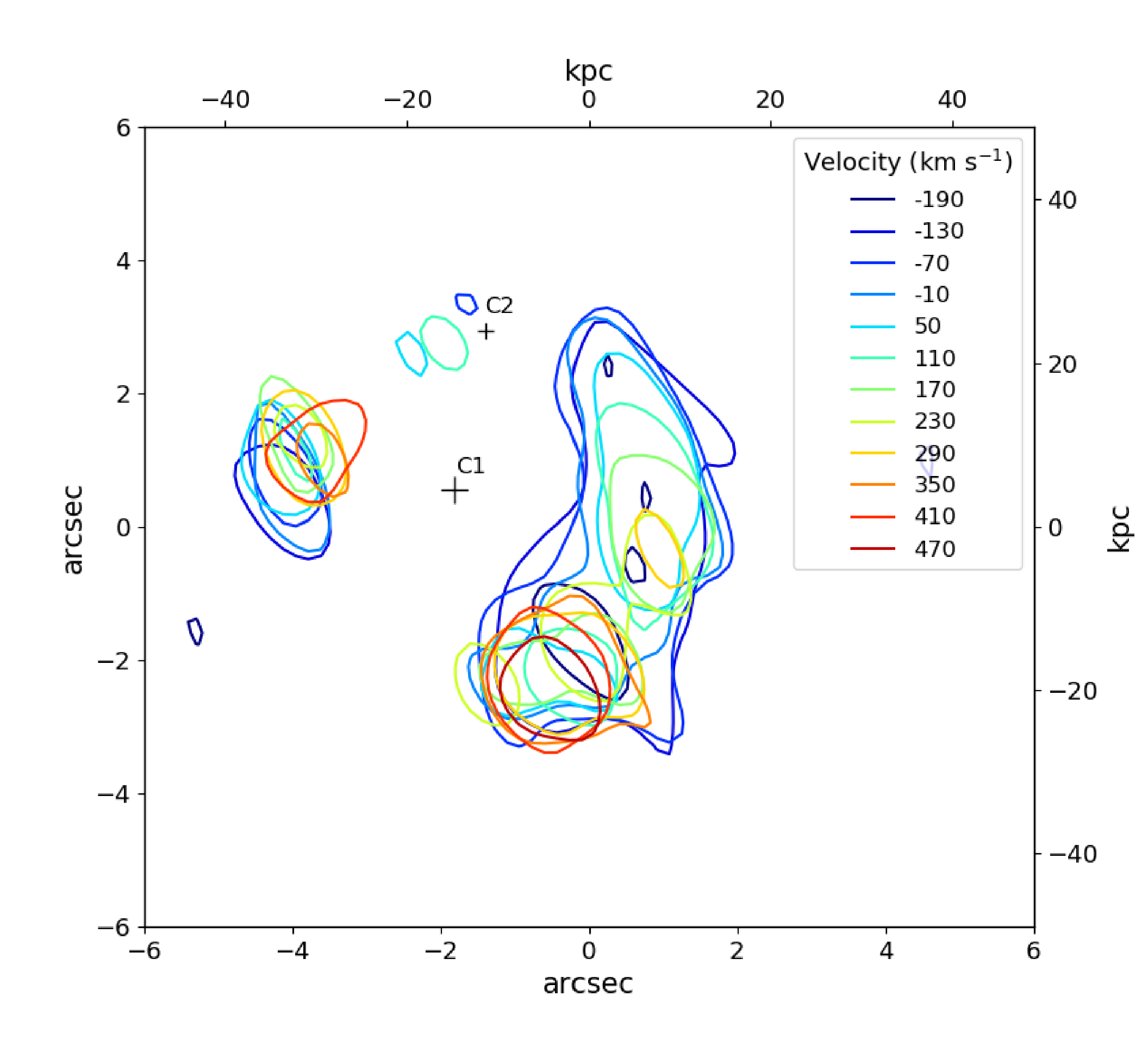}
\caption{Observed CO($J = 3-2$) channel maps of ACT\,J0209; contours 
are $4\sigma = 1.92\,{\rm mJy\,beam}^{-1}$.
Representative channels shown indicate velocities (in km\,s$^{-1}$) relative to 
$z=2.5528$. The crosses mark the positions of the two lens components. Axes indicate
angular (left, bottom) and projected linear (right, top) scales.}
\end{figure}

\begin{figure}[ht!]
\centering
\includegraphics[scale=0.38]{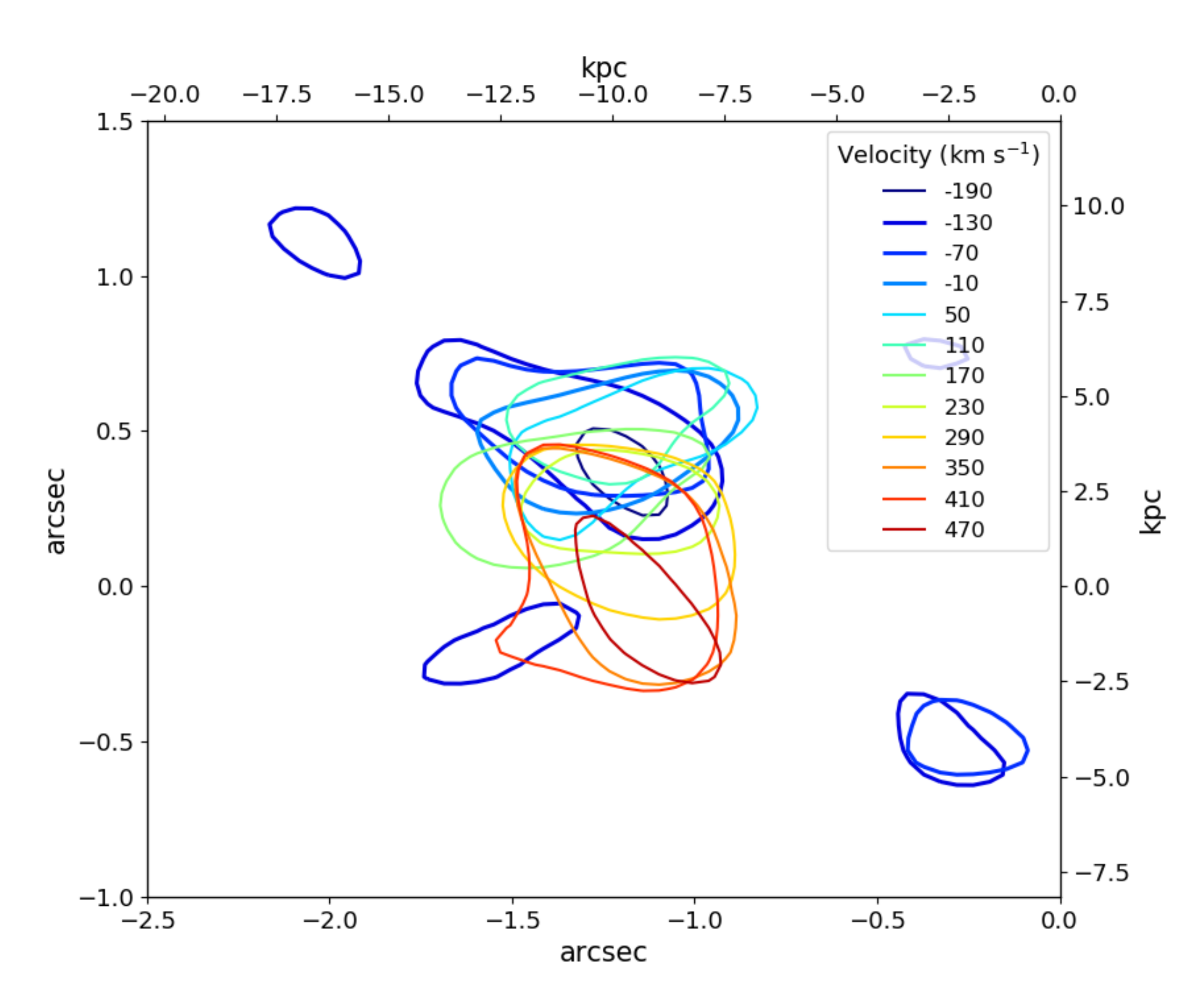}
\caption{Contours of the source-plane reconstructions for the CO($J=3-2$)
  channel maps shown in Figure 3. Source-plane reconstructions have spatially 
  varying noise and resolution, so thin (thick) contours are plotted at an
  approximate $3\sigma$ ($4\sigma$) level to indicate morphologies.  
  Representative channels shown indicate velocities (in
  km\,s$^{-1}$) relative to $z=2.5528$. Axes indicate
  angular (left, bottom) and projected linear (right, top) scales.}
\end{figure}

\begin{figure}[ht!]
\centering
\includegraphics[scale=0.36]{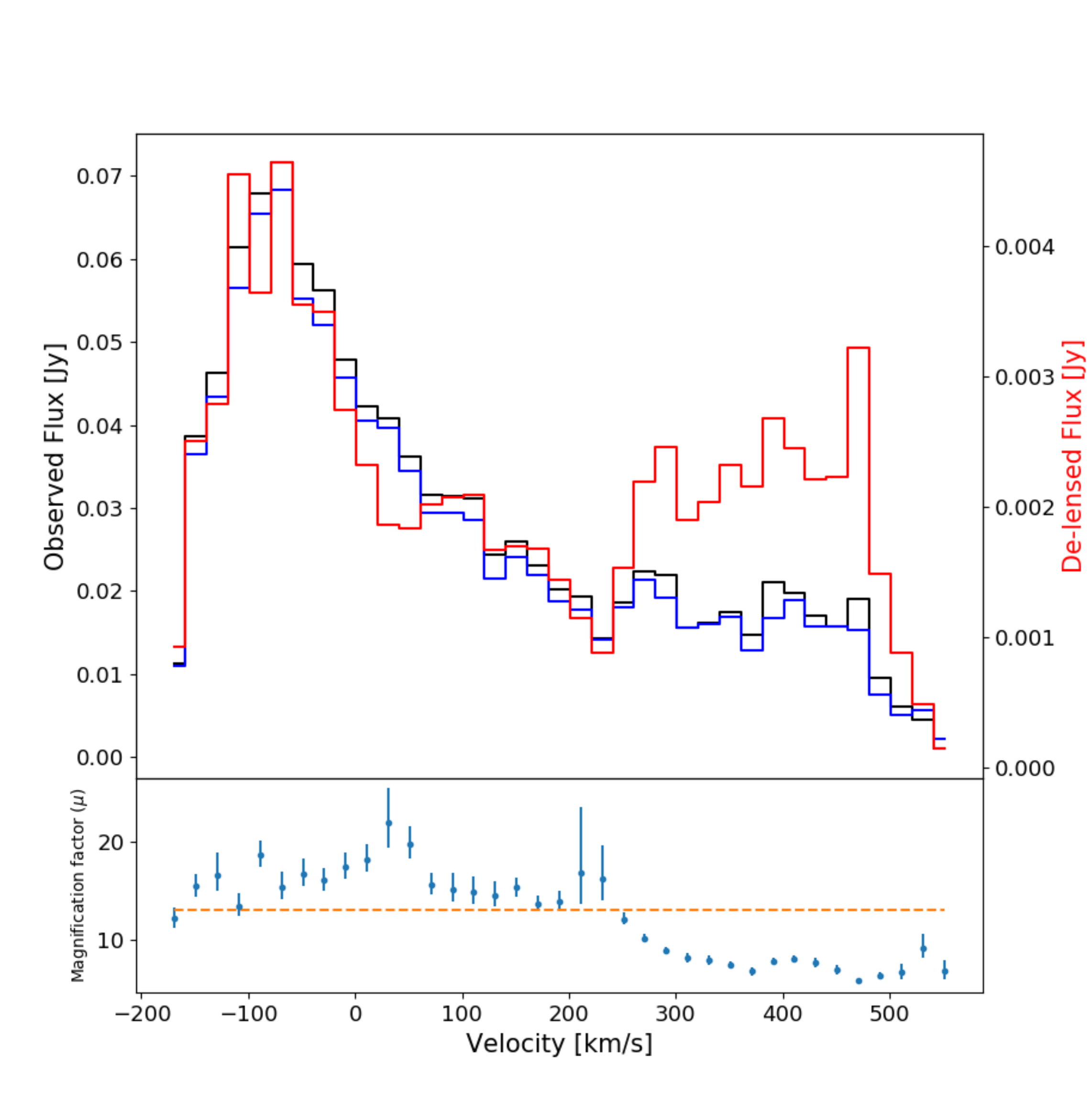}
\caption{Upper panel: observed CO($J = 3-2$) spectrum (black)
  overplotted with the spectrum of the model channel maps (blue; same scale as
  observed spectrum, at left) and
  reconstructed channel maps (red, scale at right).  
  Bluer channels (relative to $z=2.5528$) are more
  highly magnified than redder channels.  Lower panel: ratio
  between the lensed and de-lensed spectra, i.e., magnification factor on
  a per channel basis, with 16th--84th percentile uncertainties plotted as error bars.}
 \end{figure}

\section{Discussion and conclusions}

We present the first spatially resolved CO($J = 3-2$) observations of
ACT\,J0209, obtained with NOEMA, and use the pixel-based source reconstruction
technique described in \citet{tagore2014} and \citet{tagore2016}
to derive a lens model from the 37
independent velocity channels in our CO data cube.  We find that ACT\,J0209's
CO($J = 3-2$) emission is gravitationally magnified by a 
factor of $\mu \approx 7-22$ across the galaxy's velocity profile,
with a luminosity-weighted
mean $\langle \mu \rangle \approx 13$ consistent with the values found by
\citet{geach2015}. The source's unexpectedly high line brightness temperature
ratio of $r_{3,1} \approx 1.6$ is most plausibly due to external heating of
molecular clouds, although GBT pointing errors may contribute at a modest
level; a differential lensing scenario alone is
disfavored since the profiles of the GBT CO($J = 1-0$), NOEMA CO($J = 3-2$),
and ALMA CO($J = 4-3$)
spectra are broadly similar. To distinguish between these three scenarios,
high spectral and spatial resolution CO($J = 1-0$) data are needed to identify
where the line is emitted within the overall volume relative to CO($J = 3-2$).

The delensed properties of ACT\,J0209 include a CO($J = 3-2$) 
line profile that is still asymmetric across its full FWZI,
although less so than the observed spectrum; we conclude that
the distribution of molecular gas within the galaxy is not
axisymmetric.  The velocity gradient seen in the reconstructed
source-plane channel maps is, likewise, not symmetric, with
blueshifted channels especially tending to lie on top of each
other.  These asymmetries raise the question of whether we
are looking at the superposition of two galaxies (e.g., in an
early-stage merger) rather than a single rotating disk.  We
favor the single-disk interpretation for two reasons: the
very sharp drop-offs on both sides of the delensed CO($J = 3-2$) 
spectrum that closely resemble those seen for classical double-horned
disk profiles, and excitation as traced by $r_{3,1}$ (and $r_{4,1}$)
that does not vary systematically with velocity, as would be expected
for the superposition of two unmerged components \citep[e.g.,][]{sharon2015}. 
A late-stage merger containing a coalesced molecular disk, on
the other hand, remains a possibility.

\citet{geach2018} have reported an analysis of $\sim 0.25^{\prime\prime}$
resolution Atacama ALMA observations of ACT\,J0209 in the CO($J = 4-3$),
\ion{C}{1}($J = 1-0$), and CN($N = 4-3$) lines.  Those authors conclude that
the source's CO($J = 4-3$) emission arises in a rotating ring whose nearly
north/south kinematic major axis, inferred inclination ($i \approx 50^\circ$),
and dynamical mass are consistent within uncertainties with our combined
constraint on $M_{\rm dyn}\,{\rm sin}^2\,i$.  Their best estimate of the cold
gas mass from the galaxy's \ion{C}{1}($J = 1-0$) flux, however, is more than
double what is inferred from its CO($J = 1 - 0$) flux in this work.  In
contrast to our lens model, which is {\it derived} from 37 independent
CO($J = 3-2$) velocity channels, \citet{geach2018} derive their lens model
from an integrated CO($J = 4-3$) moment map before {\it applying} it to the
individual CO($J = 4-3$) velocity channels.  In this case, the loss of
information about the lensing potential that can in principle occur when a
spectral line data cube is collapsed into a moment map
\citep[see, e.g.,][]{hezaveh2013} is more than
compensated for by the superior resolution and sensitivity of the ALMA data.
Our work nevertheless demonstrates the great potential of pixel-based lens
modeling for the recovery of non-parametric structures in source-plane
channel maps, provided that (as here) observations resolve a source well but
do not resolve out significant fractions of its emission.

\section{Acknowledgments}

The authors thank Melanie Krips, Catie Raney, and Anthony Young for helpful
interactions, Min Yun for information on unpublished GBT observations, and an
anonymous referee for helpful comments.
This work was based on observations carried out under project number W16DX 
with the IRAM NOEMA Interferometer.  IRAM is supported by INSU/CNRS (France),
MPG (Germany), and IGN (Spain). 
This work was supported by the U.S. National Science Foundation through awards AST-0955810 to A.J.B., AST-1716585 to C.R.K., and AST-0408698 for the ACT project,
along with awards PHY-0855887 and PHY-1214379. Funding was also provided by
Princeton University, the University of Pennsylvania, and a Canada Foundation
for Innovation (CFI) award to the University of British Columbia. ACT operates
in the Parque Astron\'omico Atacama in northern Chile under the auspices of
the Comisi\'on Nacional de Investigaci\'on Cient\'ifica y Tecnol\'ogica de
Chile (CONICYT). C.L. thanks CONICYT for grant Anillo ACT-1417.
Computations were performed on the GPC supercomputer at the
SciNet HPC Consortium. SciNet is funded by the CFI under the auspices of
Compute Canada, the Government of Ontario, the Ontario Research FundResearch
Excellence, and the University of Toronto.

\acknowledgments

\newpage

\bibliographystyle{apj}

\begin{thebibliography}{}
\expandafter\ifx\csname natexlab\endcsname\relax\def\natexlab#1{#1}\fi

\bibitem[{{Abazajian} {et~al.}(2009){Abazajian}, {Adelman-McCarthy},
  {Ag{\"u}eros}, {Allam}, {Allende Prieto}, {An}, {Anderson}, {Anderson},
  {Annis}, {Bahcall}, \& et~al.}]{abazajian2009}
{Abazajian}, K.~N., {Adelman-McCarthy}, J.~K., {Ag{\"u}eros}, M.~A., {et~al.}
  2009, \apjs, 182, 543

\bibitem[{{Barger} {et~al.}(1998){Barger}, {Cowie}, {Sanders}, {Fulton},
  {Taniguchi}, {Sato}, {Kawara}, \& {Okuda}}]{barger98}
{Barger}, A.~J., {Cowie}, L.~L., {Sanders}, D.~B., {et~al.} 1998, \nat, 394,
  248

\bibitem[{{Becker} {et~al.}(1995){Becker}, {White}, \& {Helfand}}]{becker1995}
{Becker}, R.~H., {White}, R.~L., \& {Helfand}, D.~J. 1995, \apj, 450, 559

\bibitem[{{Blain} {et~al.}(2002){Blain}, {Smail}, {Ivison}, {Kneib}, \&
  {Frayer}}]{blain2002}
{Blain}, A.~W., {Smail}, I., {Ivison}, R.~J., {Kneib}, J.-P., \& {Frayer},
  D.~T. 2002, \physrep, 369, 111

\bibitem[{{Bolatto} {et~al.}(2013){Bolatto}, {Wolfire}, \&
  {Leroy}}]{bolatto2013}
{Bolatto}, A.~D., {Wolfire}, M., \& {Leroy}, A.~K. 2013, \araa, 51, 207

\bibitem[{{Braine} \& {Combes}(1992)}]{braine1992}
{Braine}, J., \& {Combes}, F. 1992, \aap, 264, 433

\bibitem[{{Ca{\~n}ameras} {et~al.}(2015){Ca{\~n}ameras}, {Nesvadba}, {Guery},
  {McKenzie}, {K{\"o}nig}, {Petitpas}, {Dole}, {Frye}, {Flores-Cacho},
  {Montier}, {Negrello}, {Beelen}, {Boone}, {Dicken}, {Lagache}, {Le Floc'h},
  {Altieri}, {B{\'e}thermin}, {Chary}, {de Zotti}, {Giard}, {Kneissl}, {Krips},
  {Malhotra}, {Martinache}, {Omont}, {Pointecouteau}, {Puget}, {Scott},
  {Soucail}, {Valtchanov}, {Welikala}, \& {Yan}}]{canameras2015}
{Ca{\~n}ameras}, R., {Nesvadba}, N.~P.~H., {Guery}, D., {et~al.} 2015, \aap,
  581, A105

\bibitem[{{Casey} {et~al.}(2014){Casey}, {Narayanan}, \& {Cooray}}]{casey2014}
{Casey}, C.~M., {Narayanan}, D., \& {Cooray}, A. 2014, \physrep, 541, 45

\bibitem[{{Castets} {et~al.}(1990){Castets}, {Duvert}, {Dutrey}, {Bally}, {Langer}, \& {Wilson}}]{castets1990}
{Castets}, A., {Duvert}, G., {Dutrey}, A., et~al. 1990, \aap, 234, 469

\bibitem[{{Dumke} {et~al.}(2001){Dumke}, {Nieten}, {Thuma}, {Wielebinski}, \&
  {Walsh}}]{dumke2001}
{Dumke}, M., {Nieten}, C., {Thuma}, G., {Wielebinski}, R., \& {Walsh}, W. 2001,
  \aap, 373, 853

\bibitem[{{Faber} \& {Jackson}(1976)}]{faber1976}
{Faber}, S.~M., \& {Jackson}, R.~E. 1976, \apj, 204, 668

\bibitem[{{Fowler} {et~al.}(2007){Fowler}, {Niemack}, {Dicker}, {Aboobaker},
  {Ade}, {Battistelli}, {Devlin}, {Fisher}, {Halpern}, {Hargrave}, {Hincks},
  {Kaul}, {Klein}, {Lau}, {Limon}, {Marriage}, {Mauskopf}, {Page}, {Staggs},
  {Swetz}, {Switzer}, {Thornton}, \& {Tucker}}]{fowler2007}
{Fowler}, J.~W., {Niemack}, M.~D., {Dicker}, S.~R., {et~al.} 2007, \ao, 46,
  3444

\bibitem[{{Fu} {et~al.}(2013){Fu}, {Cooray}, {Feruglio}, {Ivison}, {Riechers},
  {Gurwell}, {Bussmann}, {Harris}, {Altieri}, {Aussel}, {Baker}, {Bock},
  {Boylan-Kolchin}, {Bridge}, {Calanog}, {Casey}, {Cava}, {Chapman},
  {Clements}, {Conley}, {Cox}, {Farrah}, {Frayer}, {Hopwood}, {Jia}, {Magdis},
  {Marsden}, {Mart{\'{\i}}nez-Navajas}, {Negrello}, {Neri}, {Oliver}, {Omont},
  {Page}, {P{\'e}rez-Fournon}, {Schulz}, {Scott}, {Smith}, {Vaccari},
  {Valtchanov}, {Vieira}, {Viero}, {Wang}, {Wardlow}, \& {Zemcov}}]{fu2013}
{Fu}, H., {Cooray}, A., {Feruglio}, C., {et~al.} 2013, \nat, 498, 338

\bibitem[{{Geach} {et~al.}(2018){Geach}, {Ivison}, {Dye}, \&
  {Oteo}}]{geach2018}
{Geach}, J.~E., {Ivison}, R.~J., {Dye}, S., \& {Oteo}, I. 2018, \apj, 866, L12

\bibitem[{{Geach} {et~al.}(2015){Geach}, {More}, {Verma}, {Marshall},
  {Jackson}, {Belles}, {Beswick}, {Baeten}, {Chavez}, {Cornen}, {Cox}, {Erben},
  {Erickson}, {Garrington}, {Harrison}, {Harrington}, {Hughes}, {Ivison},
  {Jordan}, {Lin}, {Leauthaud}, {Lintott}, {Lynn}, {Kapadia}, {Kneib},
  {Macmillan}, {Makler}, {Miller}, {Monta{\~n}a}, {Mujica}, {Muxlow},
  {Narayanan}, {O'Briain}, {O'Brien}, {Oguri}, {Paget}, {Parrish}, {Ross},
  {Rozo}, {Rusu}, {Rykoff}, {Sanchez-Arg{\"u}elles}, {Simpson}, {Snyder},
  {Schloerb}, {Tecza}, {Wang}, {Van Waerbeke}, {Wilcox}, {Viero}, {Wilson},
  {Yun}, \& {Zeballos}}]{geach2015}
{Geach}, J.~E., {More}, A., {Verma}, A., {et~al.} 2015, \mnras, 452, 502

\bibitem[{{Gierens} {et~al.}(1992){Gierens}, {Stutzki}, \& {Winnewisser}}]{gierens1992} {Gierens}, K.~M., {Stutzki}, J., \& {Winnewisser}, G. 1992, \aap, 259, 271

\bibitem[{{Gildas Team}(2013)}]{gildas2013}
{Gildas Team}. 2013, {GILDAS: Grenoble Image and Line Data Analysis Software},
  Astrophysics Source Code Library, ascl:1305.010

\bibitem[{{Hainline} {et~al.}(2006){Hainline}, {Blain}, {Greve}, {Chapman}, {Smail}, \& {Ivison}}]{hainline2006} {Hainline}, L.~J., {Blain}, A.~W., {Greve}, T., et~al. 2006, \apj, 650, 614

\bibitem[{{Harrington} {et~al.}(2016){Harrington}, {Yun}, {Cybulski}, {Wilson},
  {Aretxaga}, {Chavez}, {De la Luz}, {Erickson}, {Ferrusca}, {Gallup},
  {Hughes}, {Monta{\~n}a}, {Narayanan}, {S{\'a}nchez-Arg{\"u}elles},
  {Schloerb}, {Souccar}, {Terlevich}, {Terlevich}, {Zeballos}, \&
  {Zavala}}]{harrington2016}
{Harrington}, K.~C., {Yun}, M.~S., {Cybulski}, R., {et~al.} 2016, \mnras, 458,
  4383

\bibitem[{{Harrington} {et~al.}(2018){Harrington}, {Yun}, {Magnelli}, {Frayer},
  {Karim}, {Wei\ss}, {Riechers}, {Jim\'enez-Andrade}, {Berman}, {Lowenthal},
  \& {Bertoldi}}]{harrington2018}
{Harrington}, K.~C., {Yun}, M.~S., {Magnelli}, B., {et~al.} 2018, \mnras, 474,
  3866

\bibitem[{{Harris} {et~al.}(2012){Harris}, {Baker}, {Frayer}, {Smail},
  {Swinbank}, {Riechers}, {van der Werf}, {Auld}, {Baes}, {Bussmann},
  {Buttiglione}, {Cava}, {Clements}, {Cooray}, {Dannerbauer}, {Dariush}, {De
  Zotti}, {Dunne}, {Dye}, {Eales}, {Fritz}, {Gonz{\'a}lez-Nuevo}, {Hopwood},
  {Ibar}, {Ivison}, {Jarvis}, {Maddox}, {Negrello}, {Rigby}, {Smith}, {Temi},
  \& {Wardlow}}]{harris2012}
{Harris}, A.~I., {Baker}, A.~J., {Frayer}, D.~T., {et~al.} 2012, \apj, 752, 152

\bibitem[{{Hezaveh} {et~al.}(2013){Hezaveh}, {Dalal}, {Holder}, {Kuhlen},
  {Marrone}, {Murray}, \& {Vieira}}]{hezaveh2013}
{Hezaveh}, Y., {Dalal}, N., {Holder}, G., {et~al.} 2013, \apj, 767, 9

\bibitem[{{Hughes} {et~al.}(1998){Hughes}, {Serjeant}, {Dunlop},
  {Rowan-Robinson}, {Blain}, {Mann}, {Ivison}, {Peacock}, {Efstathiou}, {Gear},
  {Oliver}, {Lawrence}, {Longair}, {Goldschmidt}, \& {Jenness}}]{hughes98}
{Hughes}, D.~H., {Serjeant}, S., {Dunlop}, J., {et~al.} 1998, \nat, 394, 241

\bibitem[{{Ivison} {et~al.}(2011){Ivison}, {Papadopoulos}, {Smail}, {Greve},
  {Thomson}, {Xilouris}, \& {Chapman}}]{ivison2011}
{Ivison}, R.~J., {Papadopoulos}, P.~P., {Smail}, I., {et~al.} 2011, \mnras,
  412, 1913

\bibitem[{{Ivison} {et~al.}(2013){Ivison}, {Swinbank}, {Smail}, {Harris},
  {Bussmann}, {Cooray}, {Cox}, {Fu}, {Kov{\'a}cs}, {Krips}, {Narayanan},
  {Negrello}, {Neri}, {Pe{\~n}arrubia}, {Richard}, {Riechers}, {Rowlands},
  {Staguhn}, {Targett}, {Amber}, {Baker}, {Bourne}, {Bertoldi}, {Bremer},
  {Calanog}, {Clements}, {Dannerbauer}, {Dariush}, {De Zotti}, {Dunne},
  {Eales}, {Farrah}, {Fleuren}, {Franceschini}, {Geach}, {George}, {Helly},
  {Hopwood}, {Ibar}, {Jarvis}, {Kneib}, {Maddox}, {Omont}, {Scott}, {Serjeant},
  {Smith}, {Thompson}, {Valiante}, {Valtchanov}, {Vieira}, \& {van der
  Werf}}]{ivison2013}
{Ivison}, R.~J., {Swinbank}, A.~M., {Smail}, I., {et~al.} 2013, \apj, 772, 137

\bibitem[{{Kramer} {et~al.}(1996){Kramer}, {Stutzki}, \& {Winnewisser}}]{kramer1996} {Kramer}, C., {Stutzki}, J., \& {Winnewisser}, G. 1996, \aap, 307, 915

\bibitem[{{Marsden} {et~al.}(2014){Marsden}, {Gralla}, {Marriage}, {Switzer},
  {Partridge}, {Massardi}, {Morales}, {Addison}, {Bond}, {Crichton}, {Das},
  {Devlin}, {D{\"u}nner}, {Hajian}, {Hilton}, {Hincks}, {Hughes}, {Irwin},
  {Kosowsky}, {Menanteau}, {Moodley}, {Niemack}, {Page}, {Reese}, {Schmitt},
  {Sehgal}, {Sievers}, {Staggs}, {Swetz}, {Thornton}, \&
  {Wollack}}]{marsden2014}
{Marsden}, D., {Gralla}, M., {Marriage}, T.~A., {et~al.} 2014, \mnras, 439,
  1556

\bibitem[{{Miller} {et~al.}(2018){Miller}, {Chapman}, {Aravena}, {Ashby},
  {Hayward}, {Vieira}, {Wei{\ss}}, {Babul}, {B{\'e}thermin}, {Bradford},
  {Brodwin}, {Carlstrom}, {Chen}, {Cunningham}, {De Breuck}, {Gonzalez},
  {Greve}, {Harnett}, {Hezaveh}, {Lacaille}, {Litke}, {Ma}, {Malkan},
  {Marrone}, {Morningstar}, {Murphy}, {Narayanan}, {Pass}, {Perry}, {Phadke},
  {Rennehan}, {Rotermund}, {Simpson}, {Spilker}, {Sreevani}, {Stark},
  {Strandet}, \& {Strom}}]{miller2018}
{Miller}, T.~B., {Chapman}, S.~C., {Aravena}, M., {et~al.} 2018, \nat, 556, 469

\bibitem[{{Mocanu} {et~al.}(2013){Mocanu}, {Crawford}, {Vieira}, {Aird},
  {Aravena}, {Austermann}, {Benson}, {B{\'e}thermin}, {Bleem}, {Bothwell},
  {Carlstrom}, {Chang}, {Chapman}, {Cho}, {Crites}, {de Haan}, {Dobbs},
  {Everett}, {George}, {Halverson}, {Harrington}, {Hezaveh}, {Holder},
  {Holzapfel}, {Hoover}, {Hrubes}, {Keisler}, {Knox}, {Lee}, {Leitch},
  {Lueker}, {Luong-Van}, {Marrone}, {McMahon}, {Mehl}, {Meyer}, {Mohr},
  {Montroy}, {Natoli}, {Padin}, {Plagge}, {Pryke}, {Rest}, {Reichardt}, {Ruhl},
  {Sayre}, {Schaffer}, {Shirokoff}, {Spieler}, {Spilker}, {Stalder},
  {Staniszewski}, {Stark}, {Story}, {Switzer}, {Vanderlinde}, \&
  {Williamson}}]{mocanu2013}
{Mocanu}, L.~M., {Crawford}, T.~M., {Vieira}, J.~D., {et~al.} 2013, \apj, 779,
  61

\bibitem[{{Nayyeri} {et~al.}(2016){Nayyeri}, {Keele}, {Cooray}, {Riechers},
  {Ivison}, {Harris}, {Frayer}, {Baker}, {Chapman}, {Eales}, {Farrah}, {Fu},
  {Marchetti}, {Marques-Chaves}, {Martinez-Navajas}, {Oliver}, {Omont},
  {Perez-Fournon}, {Scott}, {Vaccari}, {Vieira}, {Viero}, {Wang}, \&
  {Wardlow}}]{nayyeri2016}
{Nayyeri}, H., {Keele}, M., {Cooray}, A., {et~al.} 2016, \apj, 823, 17

\bibitem[{{Negrello} {et~al.}(2010){Negrello}, {Hopwood}, {De Zotti}, {Cooray},
  {Verma}, {Bock}, {Frayer}, {Gurwell}, {Omont}, {Neri}, {Dannerbauer},
  {Leeuw}, {Barton}, {Cooke}, {Kim}, {da Cunha}, {Rodighiero}, {Cox},
  {Bonfield}, {Jarvis}, {Serjeant}, {Ivison}, {Dye}, {Aretxaga}, {Hughes},
  {Ibar}, {Bertoldi}, {Valtchanov}, {Eales}, {Dunne}, {Driver}, {Auld},
  {Buttiglione}, {Cava}, {Grady}, {Clements}, {Dariush}, {Fritz}, {Hill},
  {Hornbeck}, {Kelvin}, {Lagache}, {Lopez-Caniego}, {Gonzalez-Nuevo}, {Maddox},
  {Pascale}, {Pohlen}, {Rigby}, {Robotham}, {Simpson}, {Smith}, {Temi},
  {Thompson}, {Woodgate}, {York}, {Aguirre}, {Beelen}, {Blain}, {Baker},
  {Birkinshaw}, {Blundell}, {Bradford}, {Burgarella}, {Danese}, {Dunlop},
  {Fleuren}, {Glenn}, {Harris}, {Kamenetzky}, {Lupu}, {Maddalena}, {Madore},
  {Maloney}, {Matsuhara}, {Micha{\l}owski}, {Murphy}, {Naylor}, {Nguyen},
  {Popescu}, {Rawlings}, {Rigopoulou}, {Scott}, {Scott}, {Seibert}, {Smail},
  {Tuffs}, {Vieira}, {van der Werf}, \& {Zmuidzinas}}]{negrello2010}
{Negrello}, M., {Hopwood}, R., {De Zotti}, G., {et~al.} 2010, Science, 330, 800

\bibitem[{{Peng} {et~al.}(2002){Peng}, {Ho}, {Impey}, \& {Rix}}]{peng2002}
{Peng}, C.~Y., {Ho}, L.~C., {Impey}, C.~D., \& {Rix}, H.-W. 2002, \aj, 124, 266

\bibitem[{{Perley} \& {Butler}(2013)}]{perley2013}
{Perley}, R.~A., \& {Butler}, B.~J. 2013, \apj, 204, 19

\bibitem[{{Pineda} \& {Bensch}(2007){Pineda} \& {Bensch}}]{pineda2007} {Pineda}, J.~L. \& {Bensch}, F. 2007, \aap, 470, 615

\bibitem[{{Riechers} {et~al.}(2011){Riechers}, {Carilli}, {Walter}, {Weiss},
  {Wagg}, {Bertoldi}, {Downes}, {Henkel}, \& {Hodge}}]{riechers2011a}
{Riechers}, D.~A., {Carilli}, L.~C., {Walter}, F., {et~al.} 2011, \apjl, 733,
  L11

\bibitem[{{Sharon} {et~al.}(2015){Sharon}, {Baker}, {Harris}, {Tacconi},
  {Lutz}, \& {Longmore}}]{sharon2015}
{Sharon}, C.~E., {Baker}, A.~J., {Harris}, A.~I., {et~al.} 2015, \apj, 798, 133

\bibitem[{{Sharon} {et~al.}(2016){Sharon}, {Riechers}, {Hodge}, {Carilli},
  {Walter}, {Wei{\ss}}, {Knudsen}, \& {Wagg}}]{sharon2016}
{Sharon}, C.~E., {Riechers}, D.~A., {Hodge}, J., {et~al.} 2016, \apj, 827, 18

\bibitem[{{Smail} {et~al.}(1997){Smail}, {Ivison}, \& {Blain}}]{smail97}
{Smail}, I., {Ivison}, R.~J., \& {Blain}, A.~W. 1997, \apjl, 490, L5

\bibitem[{{Su} {et~al.}(2017){Su}, {Marriage}, {Asboth}, {Baker}, {Bond},
  {Crichton}, {Devlin}, {D{\"u}nner}, {Farrah}, {Frayer}, {Gralla}, {Hall},
  {Halpern}, {Harris}, {Hilton}, {Hincks}, {Hughes}, {Niemack}, {Page},
  {Partridge}, {Rivera}, {Scott}, {Sievers}, {Thornton}, {Viero}, {Wang},
  {Wollack}, \& {Zemcov}}]{su2017}
{Su}, T., {Marriage}, T.~A., {Asboth}, V., {et~al.} 2017, \mnras, 464, 968

\bibitem[{{Swetz} {et~al.}(2011){Swetz}, {Ade}, {Amiri}, {Appel},
  {Battistelli}, {Burger}, {Chervenak}, {Devlin}, {Dicker}, {Doriese},
  {D{\"u}nner}, {Essinger-Hileman}, {Fisher}, {Fowler}, {Halpern},
  {Hasselfield}, {Hilton}, {Hincks}, {Irwin}, {Jarosik}, {Kaul}, {Klein},
  {Lau}, {Limon}, {Marriage}, {Marsden}, {Martocci}, {Mauskopf}, {Moseley},
  {Netterfield}, {Niemack}, {Nolta}, {Page}, {Parker}, {Staggs}, {Stryzak},
  {Switzer}, {Thornton}, {Tucker}, {Wollack}, \& {Zhao}}]{swetz2011}
{Swetz}, D.~S., {Ade}, P.~A.~R., {Amiri}, M., {et~al.} 2011, \apjs, 194, 41

\bibitem[{{Tagore} \& {Jackson}(2016)}]{tagore2016}
{Tagore}, A.~S., \& {Jackson}, N. 2016, \mnras, 457, 3066

\bibitem[{{Tagore} \& {Keeton}(2014)}]{tagore2014}
{Tagore}, A.~S., \& {Keeton}, C.~R. 2014, \mnras, 445, 694

\bibitem[{{Trippe} {et~al.}(2012){Trippe}, {Neri}, {Krips}, {Castro-Carrizo}, {Bremer}, {Pi\'etu}, \& {Winters}}]{trippe2012} 
{Trippe}, S., {Neri}, R., {Krips}, M. {et~al.} 2012, \aap, 540, A74

\bibitem[{{Turner} {et~al.}(1993){Turner}, {Hurt}, \& {Hudson}}]{turner1993}
{Turner}, J.~L., {Hurt}, R.~L., \& {Hudson}, D.~Y. 1993, \apjl, 413, L19

\bibitem[{{Vieira} {et~al.}(2010){Vieira}, {Crawford}, {Switzer}, {Ade},
  {Aird}, {Ashby}, {Benson}, {Bleem}, {Brodwin}, {Carlstrom}, {Chang}, {Cho},
  {Crites}, {de Haan}, {Dobbs}, {Everett}, {George}, {Gladders}, {Hall},
  {Halverson}, {High}, {Holder}, {Holzapfel}, {Hrubes}, {Joy}, {Keisler},
  {Knox}, {Lee}, {Leitch}, {Lueker}, {Marrone}, {McIntyre}, {McMahon}, {Mehl},
  {Meyer}, {Mohr}, {Montroy}, {Padin}, {Plagge}, {Pryke}, {Reichardt}, {Ruhl},
  {Schaffer}, {Shaw}, {Shirokoff}, {Spieler}, {Stalder}, {Staniszewski},
  {Stark}, {Vanderlinde}, {Walsh}, {Williamson}, {Yang}, {Zahn}, \&
  {Zenteno}}]{vieira2010}
{Vieira}, J.~D., {Crawford}, T.~M., {Switzer}, E.~R., {et~al.} 2010, \apj, 719,
  763

\bibitem[{{Wardlow} {et~al.}(2013){Wardlow}, {Cooray}, {De Bernardis},
  {Amblard}, {Arumugam}, {Aussel}, {Baker}, {B{\'e}thermin}, {Blundell},
  {Bock}, {Boselli}, {Bridge}, {Buat}, {Burgarella}, {Bussmann},
  {Cabrera-Lavers}, {Calanog}, {Carpenter}, {Casey}, {Castro-Rodr{\'{\i}}guez},
  {Cava}, {Chanial}, {Chapin}, {Chapman}, {Clements}, {Conley}, {Cox},
  {Dowell}, {Dye}, {Eales}, {Farrah}, {Ferrero}, {Franceschini}, {Frayer},
  {Frazer}, {Fu}, {Gavazzi}, {Glenn}, {Gonz{\'a}lez Solares}, {Griffin},
  {Gurwell}, {Harris}, {Hatziminaoglou}, {Hopwood}, {Hyde}, {Ibar}, {Ivison},
  {Kim}, {Lagache}, {Levenson}, {Marchetti}, {Marsden}, {Martinez-Navajas},
  {Negrello}, {Neri}, {Nguyen}, {O'Halloran}, {Oliver}, {Omont}, {Page},
  {Panuzzo}, {Papageorgiou}, {Pearson}, {P{\'e}rez-Fournon}, {Pohlen},
  {Riechers}, {Rigopoulou}, {Roseboom}, {Rowan-Robinson}, {Schulz}, {Scott},
  {Scoville}, {Seymour}, {Shupe}, {Smith}, {Streblyanska}, {Strom},
  {Symeonidis}, {Trichas}, {Vaccari}, {Vieira}, {Viero}, {Wang}, {Xu}, {Yan},
  \& {Zemcov}}]{wardlow2013}
{Wardlow}, J.~L., {Cooray}, A., {De Bernardis}, F., {et~al.} 2013, \apj, 762,
  59

\bibitem[{{Young} \& {Scoville}(1984)}]{young1984}
{Young}, J.~S., \& {Scoville}, N.~Z. 1984, \apj, 287, 153

\end{thebibliography}

\end{document}